\documentclass[a4paper,11pt]{article}%
\usepackage{graphicx}
\usepackage{amsbsy}
\usepackage{amsmath}
\usepackage{cite}
\usepackage{slashed}
\usepackage{epsfig}
\usepackage{hyperref}%
\usepackage{amsfonts}%
\usepackage{amssymb}
\hypersetup{
    colorlinks   = true,
    linkcolor    = blue,
    citecolor    = blue,
}

\begin{document}
\thispagestyle{empty}  \setcounter{page}{0}  \begin{flushright}%

LPSC15370\\
December 2015\\
\end{flushright}

\vskip           2.1 true cm

\begin{center}
{\huge Could the width of the diphoton anomaly signal\\[0pt]a three-body
decay?}\\[1.9cm]

\textsc{J\'er\'emy Bernon}$^{1}$ \textsc{and Christopher Smith}$^{2}$%
\vspace{0.5cm}\\[9pt]\smallskip{\small \textsl{\textit{Laboratoire de Physique
Subatomique et de Cosmologie, }}}\linebreak
{\small \textsl{\textit{Universit\'{e} Grenoble-Alpes, CNRS/IN2P3, 53 avenue
des Martyrs, 38026 Grenoble Cedex, France}.}} \\[1.9cm]\textbf{Abstract}\smallskip
\end{center}

\begin{quote}
\noindent The recently observed diphoton anomaly at the LHC appears to suggest
the presence of a rather broad resonance. In this note, it is pointed out that
this broadness is not called for if the two photons are produced along with an
extra state. Specifically, the diphoton invariant mass arising from various
$A\rightarrow B\gamma\gamma$ processes, with $A,B$ being scalars, fermions, or
vectors, though peaked at a rather large value, would naturally be broad and
could fit rather well the observed deviations. This interpretation has a
number of advantages over the two-photon resonance hypothesis, for example
with respect to the compatibility with the 8 TeV diphoton, dilepton or dijet
searches, and opens many new routes for New Physics model construction.

\let           \thefootnote            \relax         \footnotetext{$^{1}%
\;$bernon@lpsc.in2p3.fr} \footnotetext{$^{2}\;$chsmith@lpsc.in2p3.fr}
\end{quote}

\newpage

\section{Introduction and set-up}

Recently, a small deviation in the diphoton mass spectrum was announced by
both ATLAS~\cite{ATLAS} and CMS~\cite{CMS} at a mass of around 750 GeV. While
the statistical significance of this signal is still low, the simultaneous
observation by both experiments lends some credence to the presence of a yet
unknown resonance in this channel, and has led to an incredibly intense
phenomenological activity (see Refs.~\cite{Harigaya:2015ezk}
to~\cite{Matsuzaki:2015che}).

In this note, we want to point out that one feature of this $\gamma\gamma$
signal, namely its width, could be well explained if it arises from a
three-body decay $A\rightarrow B\gamma\gamma$, with the mass splitting
$M_{A}-M_{B}$ a bit higher than $750$ GeV. The $B$ particle would either be
stable and escape undetected, or would be produced on-shell and would
subsequently decay into some other invisible states.

Let us recall that the differential rate for the decay $A\rightarrow
B\gamma\gamma$ depends only on the invariant mass of the two photons, $z\equiv
m_{\gamma\gamma}^{2}/M_{A}^{2}$, or equivalently, on the $B$ momentum
$P_{B}\equiv|\mathbf{p}_{B}|/M_{A}=\sqrt{\lambda}/2$, with $\lambda
\equiv\lambda(1,z,r^{2})=1+z^{2}+r^{4}-2z-2r^{2}-2zr^{2}$ the standard
kinematical function and $r\equiv M_{B}/M_{A}$. Specifically,
\begin{equation}
\Gamma(A\rightarrow B\gamma\gamma)=\int_{0}^{\left(  1-r\right)  ^{2}%
}dz\,\frac{d\Gamma}{dz}[z]=\int_{0}^{(1-r^{2})/2}\frac{2P_{B}dP_{B}}%
{\sqrt{r^{2}+P_{B}^{2}}}\frac{d\Gamma}{dz}\left[  z(P_{B})=1+r^{2}%
-2\sqrt{P_{B}^{2}+r^{2}}\right]  \;.
\end{equation}
To match the observed ATLAS spectrum~\cite{ATLAS}, all that is needed is a differential
rate falling down sufficiently fast above $750$ GeV. Far below the peak, the
SM background quickly increases and would wipe out any sensitivity to the
$A\rightarrow B\gamma\gamma$ process. Still, slightly below the peak, at
around $600$ GeV, the event rate matches the background. Even if this
corresponds only to a few data point, for which the uncertainty is still
rather large, the differential rate should preferably fall down not too slowly
as $m_{\gamma\gamma}^{2}$ decreases.

\section{Effective four-point interactions}

To check whether a peaked behavior for the diphoton invariant mass spectrum is
realistic, and since the nature of the decaying state is no longer
constrained, we can consider various assignments for $A$ and $B$. Our basic
assumption is that $A$ and $B$ are neutral under the SM gauge group, but
nevertheless share some conserved charge $\chi$. If $\chi(A)=-\chi(B)$, the
effective interactions involving a pair of photons can derive from either
\begin{subequations}
\label{All2G}%
\begin{align}
\text{Scalar case}  &  :\mathcal{L}_{eff}=\frac{1}{\Lambda^{2}}(S_{A}S_{B}F_{\mu
\nu}F^{\mu\nu}+S_{A}S_{B}F_{\mu\nu}\tilde{F}^{\mu\nu})\;,\\
\text{Fermion case}  &  :\mathcal{L}_{eff}=\frac{1}{\Lambda^{3}}(\bar{\psi}_{A}%
^{C}\psi_{B}F_{\mu\nu}F^{\mu\nu}+\bar{\psi}_{A}^{C}\gamma_{5}\psi_{B}F_{\mu
\nu}\tilde{F}^{\mu\nu}+h.c.)\;,\\
\text{Vector case}  &  :\mathcal{L}_{eff}=\frac{1}{\Lambda^{4}}(A_{\alpha\beta
}B^{\alpha\beta}F_{\mu\nu}F^{\mu\nu}+A_{\alpha\beta}\tilde{B}^{\alpha\beta
}F_{\mu\nu}\tilde{F}^{\mu\nu}+...)\;,
\end{align}
where $\tilde{F}^{\mu\nu}=\frac{1}{2}\varepsilon^{\mu\nu\rho\sigma}%
F_{\rho\sigma}$ and possible Wilson coefficients dressing each operator can be
thought as being absorbed in the scale $\Lambda$ for notation clarity. These
effective operators are all independent, and assumed valid above the
electroweak scale. In this respect, they should thus actually be written in
terms of the $SU(2)_{L}$ and/or $U(1)_{Y}$ field strengths. For instance,
replacing%
\end{subequations}
\begin{equation}
F^{\mu\nu}\rightarrow B^{\mu\nu}=\cos\theta_{W}F^{\mu\nu}-\sin\theta_{W}%
Z^{\mu\nu}\;,
\end{equation}
the $\gamma\gamma$, $Z\gamma$, and $ZZ$ modes would be produced in the ratio
$1:2\tan^{2}\theta_{W}:\tan^{4}\theta_{W}$, up to kinematical effects (in
exactly the same way as for the two-body interpretation of the diphoton
anomaly, see \textit{e.g.} Ref.~\cite{Franceschini:2015kwy}). Finally, the CP symmetry
can be enforced without loss of generality, since it is always possible to set
the two photons in the adequate CP state ($F_{\mu\nu}F^{\mu\nu}$ and
$F_{\mu\nu}\tilde{F}^{\mu\nu}$ have opposite parity).

\begin{figure}[t]
\centering     \includegraphics[width=0.5\textwidth]{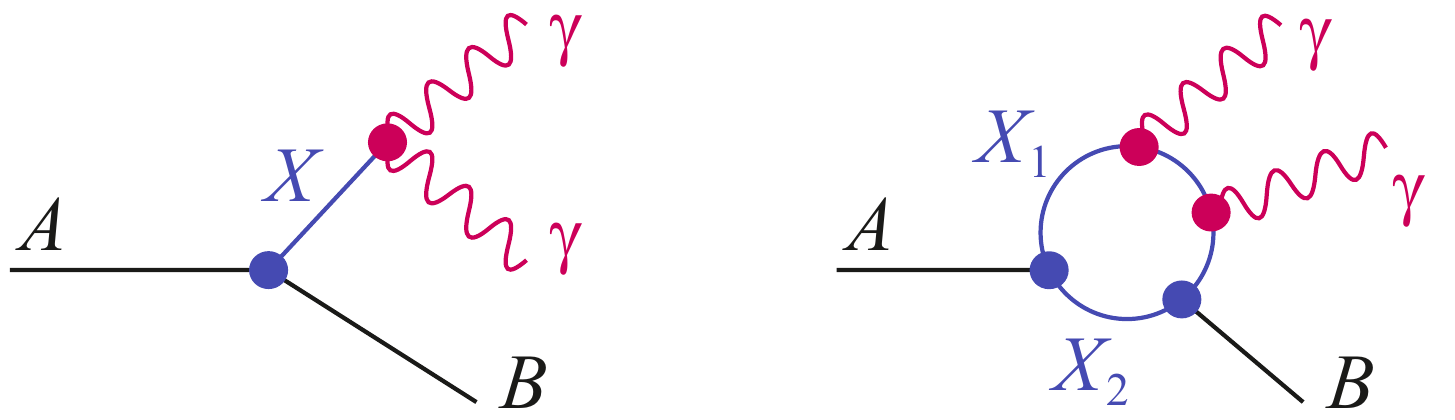}
\caption{Example of short-distance processes leading to the effective
interactions in Eq.~(\ref{All2G}). For the tree-level diagram, $X$ is a scalar
or tensor state, whose coupling to two photons must involve yet another state.
For the loop diagram, there must be a pair of states circulating the loop to
ensure $\chi$ conservation. }%
\label{Fig0}
\end{figure}

At this stage, the main issue is whether simpler interactions, as for instance those
involving a single photon, are possible. Though a full answer to this question
would require constructing full-fledged UV completions, which is well beyond our current
scope, we can nevertheless draw a number of observations. These effective
interactions could either arise at tree level or at loop level, see
Fig.~\ref{Fig0}, and in general require more than one extra state. For
instance, in the former case, the additional resonance $X$ would be a scalar
or tensor state coupled to two photons. We only allow it to be off-shell,
since otherwise the three-body signature would be lost. The $X$ would simply
be a true diphoton resonance with a mass of 750 GeV. Still, even if off-shell,
this state can couple to two photons only through additional new degrees of
freedom, for example a vector fermion loop. The main interest of this scenario is
that the single-photon processes are automatically absent.

If generated at loop level, two new states are also required in general to
ensure the conservation of $\chi$ and prevent $A,B\rightarrow\gamma\gamma$.
Both of them could be fermions when $A$ and $B$ are scalars or vectors, but at
least one new scalar or vector is needed to induce $\psi_{A}\rightarrow
\psi_{B}\gamma\gamma$. The only exception is the charged scalar loop when
$A,B$ are themselves also scalars, with a renormalizable $ABX^{+}X^{-}$
vertex. Anyway, looking at Fig.~\ref{Fig0}, it seems obvious that such loops
induce also single photon modes (along with potentially large mixings between
the two states, which we assumed have been dealt with properly so that states
occurring in the effective interactions are true mass eigenstates). Whether
such processes truly occur, and in case they do, the relative strength of the
one and two photon modes, depends on the nature of $A$ and $B$, so we now
discuss the various assignments separately.

\subsubsection*{Scalar transitions}

The single photon production $S_{A}\rightarrow S_{B}\gamma$ is forbidden by
Lorentz and gauge invariance (for the same reason as, \textit{e.g.}, $K^{+}%
\nrightarrow\pi^{+}\gamma$ or $\eta\nrightarrow\pi^{0}\gamma$). At the
renormalizable level, trivially, a direct coupling of the photon field
$A^{\mu}$ to the scalar current $S_{A}\partial_{\mu}S_{B}-S_B\partial_{\mu}%
S_{A}$ is not gauge invariant since the current is not conserved when
$m_{A}\neq m_{B}$. Beyond leading order, effective operators involving a
single photon field can be constructed, for instance
\begin{equation}
\frac{1}{\Lambda^{2}}S_{A}\partial^{\nu}S_{B}\partial^{\mu}F_{\mu\nu}\;,
\label{Scalar1G}%
\end{equation}
but the amplitude necessarily vanishes for an on-shell photon. There is no
corresponding operator involving $\tilde{F}^{\mu\nu}$, as can be easily
understood at the Feynman rule level since there are only three independent
four-vectors to be contracted with $\varepsilon^{\mu\nu\rho\sigma}$. This
implies that if $S_{A}$ and $S_{B}$ are real fields with the same parity, then
$S_{A}\rightarrow S_{B}\gamma^{\ast}$ is CP-violating (as is \textit{e.g.}
$\eta\rightarrow\pi^{0}\ell^{+}\ell^{-}$).

Interestingly, this could suffice to reduce the $S_{A}\rightarrow S_{B}%
\ell^{+}\ell^{-}$ or $S_{A}\rightarrow S_{B}q\bar{q}$ signals, even when CP
conserving. Since the effective interaction is of the same dimension as the
two-photon ones, producing the fermion pair through $A\rightarrow
B[\gamma^{\ast}\rightarrow f\bar{f}]$ is at best comparable to the
$\gamma\gamma$ mode, and could actually end up very suppressed if the
situation for $K^{0}\rightarrow\pi^{0}\gamma\gamma$ compared to $K_{S}%
\rightarrow\pi^{0}\ell^{+}\ell^{-}$ is~of~any guide~\cite{PDG}.

Coming back to the vector fermion loop, it is easy to see that it never
induces the operator Eq.~(\ref{Scalar1G}). If both scalars couple as
$S_{A,B}\bar{\psi}_{F}\psi_{F}$ or $S_{A,B}\bar{\psi}_{F}\gamma_{5}\psi_{F}$
with $\psi_{F}$ the electrically charged heavy vector fermion circulating in
the loop, then the process is CP-violating and the sum of the two diagrams
where $\psi_{F}$ circles clockwise and anticlockwise cancel each other. If one
scalar couples through $\bar{\psi}_{F}\psi_{F}$ and the other though $\bar
{\psi}_{F}\gamma_{5}\psi_{F}$, then both amplitudes are proportional to
$\varepsilon_{\mu\nu\rho\sigma}(\varepsilon_{\gamma}^{\ast})^{\mu}p_{A}^{\nu
}p_{B}^{\rho}p_{\gamma}^{\sigma}=0$ since $p_{A}=p_{B}+p_{\gamma}$. At this
level, single photon processes cannot be induced.

\subsubsection*{Vector transitions}

For the vector case, first remark that we do not consider all possible index
contractions among the four field strengths in Eq.~(\ref{All2G}), but only
some representative examples from the point of view of the differential rate.
More importantly, we have not included dimension-six operators like
$A_{\alpha}B^{\alpha}F_{\mu\nu}F^{\mu\nu}$ for three reasons. First, those
would lead to differential rates very similar to the scalar case. Second, they
may be quite complicated to generate from some UV completion. Finally, nothing
would prevent a renormalizable coupling to a single photon like $A^{\mu}%
B^{\nu}F_{\mu\nu}$. The Landau-Yang theorem does not apply without gauge
invariance or with two different vector bosons in the final state.

Even if we insist on constructing only operators involving field strengths,
the $V_{A}\rightarrow V_{B}\gamma$ process is not manifestly forbidden because
$m_{A,B}\neq0$, as can be seen starting with%
\begin{equation}
\frac{1}{\Lambda^{2}}(A_{\nu\alpha}B^{\alpha\mu}F_{\mu}^{\nu}+A_{\nu\alpha
}B^{\alpha\mu}\tilde{F}_{\mu}^{\nu}+...)\;.
\end{equation}
Nevertheless, the $V_{A}\rightarrow V_{B}\gamma$ process along with
$V_{A}\rightarrow V_{B}[\gamma^{\ast}\rightarrow\ell^{+}\ell^{-},q\bar{q}]$
could be very suppressed. Taking again the vector fermion loop, and assuming
$V_{A}$ and $V_{B}$ have both either vector or axial-vector couplings to
$\psi_{F}$, charge conjugation ensures the cancellation of all the diagrams to
which an odd number of vector fields are attached. So, instead of the
Landau-Yang theorem, what really matters in this case is the Furry theorem of
QED. Note that axial-vector couplings seem more tenable to prevent the kinetic
mixing $V_{A,B}\leftrightarrow\gamma$, though we have not analyzed the vector
coupling scenario further.

\subsubsection*{Fermion transitions}

For the fermion case, operators involving a single field strength are not
forbidden. Gauge invariance prevents the direct coupling to the fermion
current $\bar{\psi}_{A}^{C}\gamma_{\mu}\psi_{B}$, but one can construct
\begin{equation}
\frac{1}{\Lambda}(\bar{\psi}_{A}^{C}\sigma_{\mu\nu}\psi_{B}F^{\mu\nu}%
+\bar{\psi}_{A}^{C}\sigma_{\mu\nu}\psi_{B}\tilde{F}^{\mu\nu})\;.
\end{equation}
Contrary to the scalar case, these operators produce an on-shell photon, are
of lower dimension than those in Eq.~(\ref{All2G}), and both $F^{\mu\nu}$ and
$\tilde{F}^{\mu\nu}$ can occur so CP can be of no help. If arising at loop
level, there does not seem to be any obvious way to enhance the two-photon
relative to the one-photon emission (besides asking for $\psi_{B}$ to decay
rather quickly into a photon plus yet another fermion $\psi_{C}$).
Phenomenologically, the fermionic scenario is most likely to make sense only
in the tree-level hypothesis.

\section{Differential rates and interpretation}

\begin{figure}[t]
\centering        \includegraphics[width=1.\textwidth]{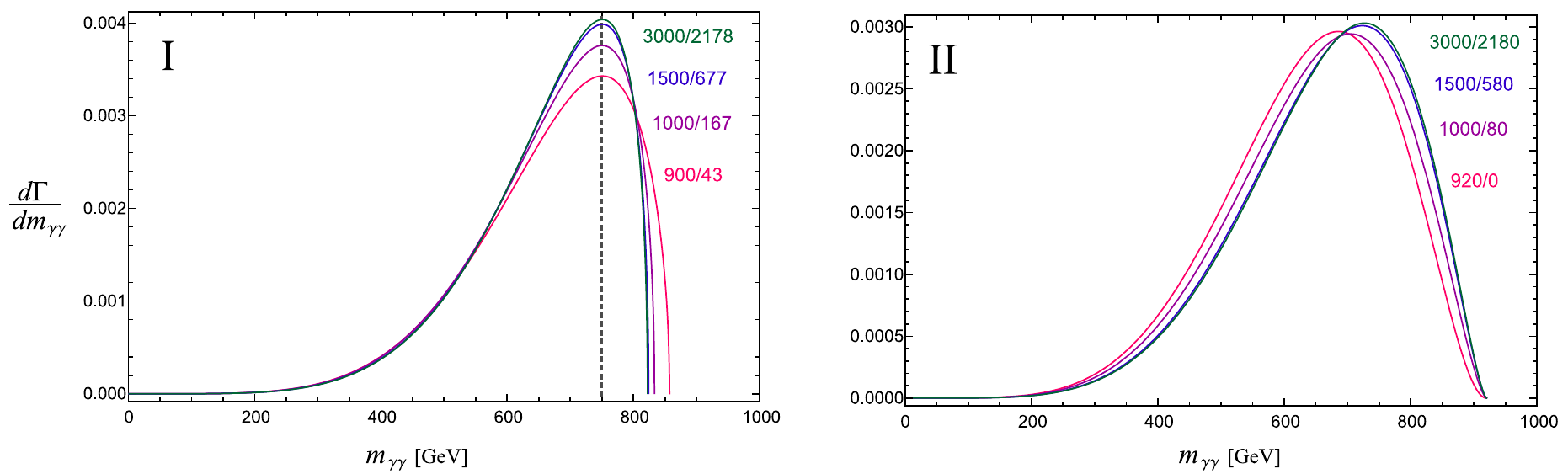}  \caption{No
matter the effective operator, differential diphoton spectra fall in one of
two classes: those peaking at high diphoton invariant mass (left panel), and
those peaking closer to the average diphoton mass (right panel). Specifically,
all but the $\bar{\psi}_{A}\gamma_{5}\psi_{B}F_{\mu\nu}\tilde{F}^{\mu\nu}$ and
$A_{\alpha\beta}\tilde{B}^{\alpha\beta}F_{\mu\nu}\tilde{F}^{\mu\nu}$ are in
the first class. Therefore, in the left panel, we show the differential rate
for the scalar case for various choices of $M_{A}$, and with $M_{B}$ fixed
(the labels close to each curve denote $M_{A}/M_{B}$, in GeV) so that the peak
is precisely at 750 GeV.}%
\label{diff_rate}%
\end{figure}

The differential rates are straightforward to compute for these various cases,
giving
\begin{subequations}
\label{scalar}%
\begin{align}
\frac{d\Gamma}{dz}(S_{A} &  \rightarrow S_{B}\gamma\gamma)_{++,-+}%
=\frac{M_{A}^{5}}{32\pi^{3}\Lambda^{4}}z^{2}\lambda(1,z,r^{2})^{1/2}\;,\\
\frac{d\Gamma}{dz}(\psi_{A} &  \rightarrow\psi_{B}\gamma\gamma)_{++}%
=\frac{M_{A}^{7}}{64\pi^{3}\Lambda^{6}}z^{2}\lambda(1,z,r^{2})^{1/2}%
((r+1)^{2}-z)\;,\\
\frac{d\Gamma}{dz}(\psi_{A} &  \rightarrow\psi_{B}\gamma\gamma)_{-+}%
=\frac{M_{A}^{7}}{16\pi^{3}\Lambda^{6}}z^{2}\lambda(1,z,r^{2})^{1/2}%
((r-1)^{2}-z)\;,\\
\frac{d\Gamma}{dz}(V_{A} &  \rightarrow V_{B}\gamma\gamma)_{++}=\frac{M_{A}%
^{9}}{4\pi^{3}\Lambda^{8}}z^{2}\lambda(1,z,r^{2})^{1/2}(6r^{2}+\lambda
(1,z,r^{2}))\;,\\
\frac{d\Gamma}{dz}(V_{A} &  \rightarrow V_{B}\gamma\gamma)_{-+}=\frac{M_{A}%
^{9}}{4\pi^{3}\Lambda^{8}}z^{2}\lambda(1,z,r^{2})^{3/2}\;,
\end{align}
where the subscripts denote a scalar (++) or pseudoscalar (-+) coupling to
$F_{\mu\nu}F^{\mu\nu}$ or $F_{\mu\nu}\tilde{F}^{\mu\nu}$, respectively. From
these shapes, we can draw a number of conclusions:

\end{subequations}
\begin{enumerate}
\item All the differential rates show a strong dependence on $m_{\gamma\gamma
}$, which can be traced to the photon momenta arising from the derivatives
present in $F_{\mu\nu}F^{\mu\nu}$ or $F_{\mu\nu}\tilde{F}^{\mu\nu}$. More
generally and model-independently, Low's theorem~\cite{Low} tells us that when
$A$ and $B$ are electrically neutral, the $A\rightarrow B\gamma\gamma$
amplitude must be at least linear in the photon energy $E_{\gamma}$ as
$E_{\gamma}\rightarrow0$. At larger $m_{\gamma\gamma}$, the squared amplitude
is dampened by the kinematical factors forcing $d\Gamma/dm_{\gamma\gamma}$ to
go back to zero at the high-energy end-point. In the middle, the differential
rate thus always shows a peak. Interestingly, requiring it to be close to its
high-energy end-point does not suffices to discriminate between spin $0$,
$1/2$, or $1$ resonances. All we can say asking for a high $m_{\gamma\gamma
}^{2}$ peak is that a few couplings cannot match the observed anomaly, with
$\bar{\psi}_{A}^{C}\gamma_{5}\psi_{B}F_{\mu\nu}\tilde{F}^{\mu\nu}$ and
$A_{\alpha\beta}\tilde{B}^{\alpha\beta}F_{\mu\nu}\tilde{F}^{\mu\nu}$
producing only a broad bump in the middle of the kinematical range, see
Fig.~\ref{diff_rate}. On the other hand, for all the other operators, the
spectrum is very similar and peaks at high diphoton invariant mass. Its shape
is quite consistent with the observed events, see Fig.~\ref{spectrum}. In this
respect, note how the peak initially gets more pronounced as $M_{A}$
increases, but quickly reaches a limiting shape and does not change
significantly beyond $M_{A}\approx1.5$ TeV.

\begin{figure}[t]
\centering        \includegraphics[width=.68\textwidth]{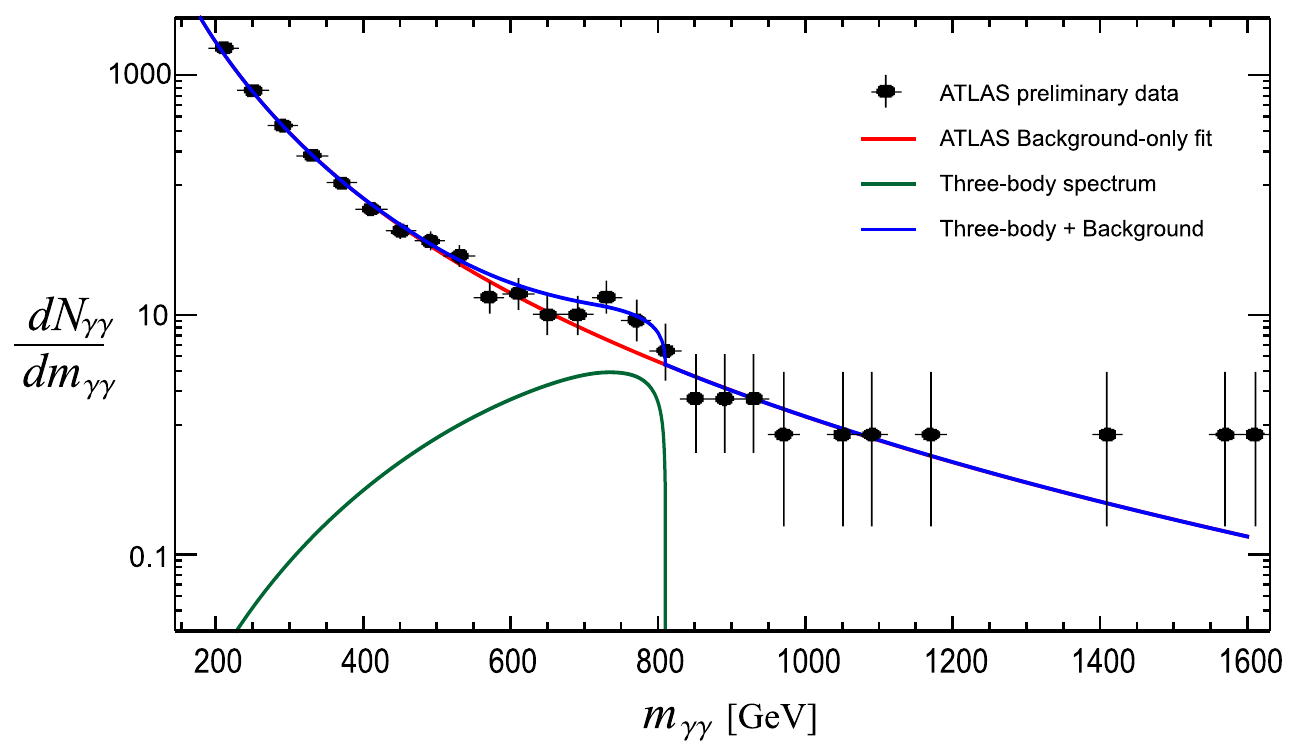}
\caption{Impact of a three-body production of a pair of photons, for
$M_{A}=1.5$~TeV and $M_{A}-M_{B}=810$~GeV, over the background observed by
ATLAS \cite{ATLAS}. Lacking all the details about the data points, their
errors, and correlations, the $A\rightarrow B\gamma\gamma$ rate is adjusted by
hand.}%
\label{spectrum}
\end{figure}

\item The mass scale of the process is always higher than $750$ GeV, because
for $m_{\gamma\gamma}$ to peak there, the mass of the decaying resonance has
to be above about $900$ GeV. Actually, it is even possible for the $A$
resonance to be well above the TeV scale. This automatically helps explaining
why no such signal was seen at $8$ TeV. Indeed, the gain factor in cross
section going from $8$~TeV to $13$~TeV, for a typical partonic production,
increases with the resonance mass. For example, if produced through the
gluon-gluon channel, the gain factor is of about $5.3$ for $M_{A}=900$~GeV,
and already nearly twice as large, $9.3$, for $M_{A}=1.5$~TeV. Note, finally,
that this also helps to pass the bounds obtained at $8$~TeV in the $\gamma
Z$~\cite{Aad:2014fha} and $ZZ$~\cite{Aad:2015kna} channels.

\item If not forbidden (see previous section), this peaked behavior of the
differential rate is not necessarily matched by single photon emission. For
example, starting with the operator in Eq.~(\ref{Scalar1G}) and coupling the
virtual photon to a Dalitz pair, we find
\begin{equation}
\frac{d\Gamma}{dz}(S_{A}\rightarrow S_{B}[\gamma^{\ast}\rightarrow f\bar
{f}])=\frac{M_{A}^{5}}{256\pi^{3}\Lambda^{4}}\lambda(1,z,r^{2})^{3/2}\;,
\end{equation}
in the limit $m_{f}\rightarrow0$, where $z=m_{\ell\ell}^{2}/M_{A}^{2}$ is now
the reduced dilepton invariant mass. In this case, the photon momentum
dependence of the effective vertex Eq.~(\ref{Scalar1G}) is compensated by the
$1/m_{\ell\ell}^{2}$ coming from the virtual photon propagator, and the
differential rate ends up maximal at $z=0$, falling off roughly linearly
towards zero as $z\rightarrow(1-r)^{2}$. With such a shape devoid of any
particular feature, and with the rather suppressed rate, the $8$~TeV bounds
are easily satisfied~\cite{Aad:2014cka} and it is not even clear such a signal could
be easily evidenced in the future.

\item Thanks to the strong peak at high $m_{\gamma\gamma}^{2}$, the invisible
state escaping the detector would carry away a moderate amount of missing
energy, as shown in Fig.~\ref{B_momentum}. Typically, the $B$ momentum in the
$A$ rest-frame peaks at about $15-20\%$ of $M_{A}$. For example, with
$M_{A}=1.5~$TeV, it is maximum for $|\mathbf{p}_{B}|\approx280$ GeV. Still,
together with observing an excess in $\gamma\gamma$ events for lower invariant
mass, it could help identify the three body nature of the process.
Alternatively, but at the cost of allowing for the new charge $\chi$ to be
violated at some point, the $B$ particle could also decay, for example into a
pair of rather soft photons or leptons which would have been cut away in
selecting the $\gamma\gamma$ candidate events. Provided this state can only be
produced via the decay of $A$, it could well have escape detection up to now.

\begin{figure}[t]
\centering        \hspace{-1.7cm}
\includegraphics[width=.55\textwidth]{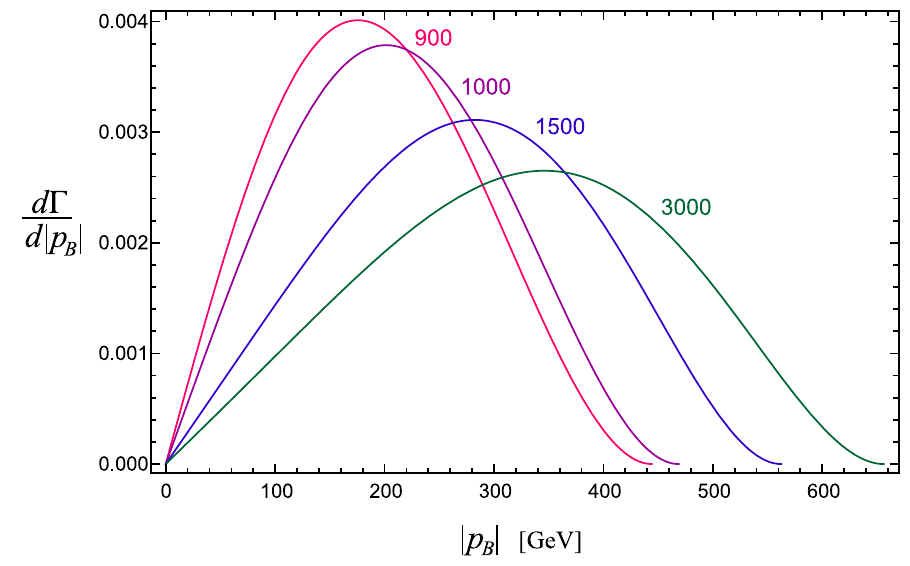}  \caption{Spectrum of the $B$
momentum in the $A$ rest-frame for various $M_{A}$ hypotheses. The effective
$S_{A}S_{B}F_{\mu\nu}F^{\mu\nu}$ operator is chosen for definiteness.}%
\label{B_momentum}%
\end{figure}

\item Finally, one could think of pushing the reasoning one step further and
consider four-body decays $A\rightarrow B+C+\gamma\gamma$. As for the
three-body processes, the momentum dependence hidden in the $F_{\mu\nu}%
F^{\mu\nu}$ or $F_{\mu\nu}\tilde{F}^{\mu\nu}$ structures still favors the
presence of a peak for rather large diphoton invariant mass. On the other
hand, a four-body interpretation has several short-comings. It is less trivial
to design a single conserved charge able to prevent simpler cascade decays, or
involving only one photon. In addition, the dimension of the effective
operators are larger and the rate further phase-space suppressed, so $\Lambda$
has to be systematically lower casting serious doubts on the effective
treatment. Moreover, the missing energy carried away by both $B$ and $C$ could
be too large to have stayed unnoticed.
\end{enumerate}

\section{On the scale of the effective operators}

To explain the rather large observed $\gamma\gamma$ anomaly, the decay width
into two photons and the production of the $A$ resonance have to be
sufficiently large. Up to now we did not consider the latter production, since
our goal is to show the compatibility of the three-body hypothesis with the
shape of the diphoton anomaly. Also, dealing with both production and decay is
necessarily more model-dependent. So in this section, we will present a few
arguments and, staying as generic as possible, give some estimates of the
scale of the effective interactions.

As a first handle, we consider the scalar case with the production mechanism
$gg\rightarrow S_{A}$:
\begin{equation}
\mathcal{L}_{eff}=\frac{\kappa_{gg}}{\Lambda}S_{A}G_{\mu\nu}G^{\mu\nu
}+\frac{\kappa_{\gamma\gamma}}{\Lambda^{2}}S_{A}S_{B}F_{\mu\nu}F^{\mu\nu}.
\end{equation}

\begin{figure}[t]
\centering        \includegraphics[width=.65\textwidth]{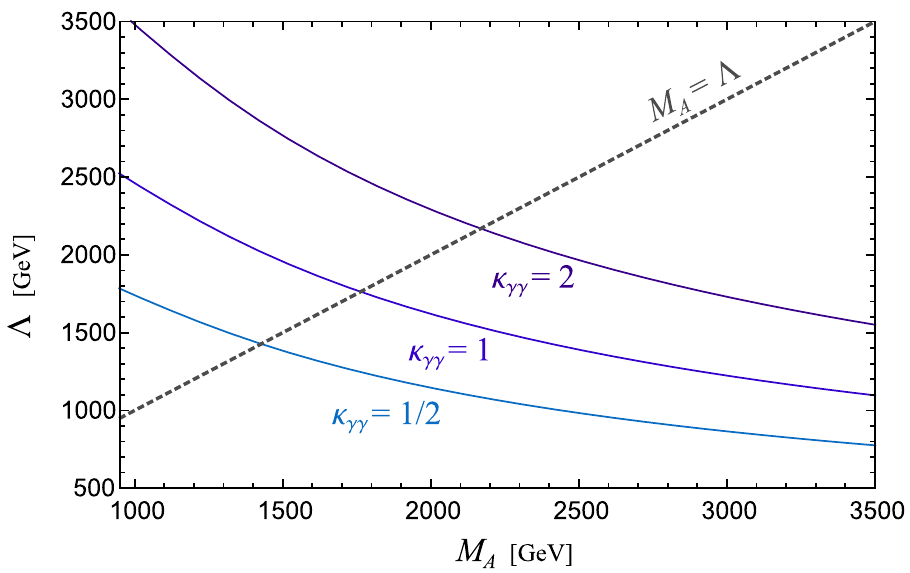}
\caption{Scale $\Lambda$ as a function of $M_{A}$, setting $\kappa
_{\gamma\gamma}=1/2,1,2$, required to reproduce the observed diphoton
production rate. For simplicity, following Ref.~\cite{Franceschini:2015kwy},
we assume for this plot that the electromagnetic width is sub-leading compared
to the $gg$ channel and neglect the evolution of the partonic $gg$
density as a function of $M_A$, so as to fix $\Gamma(A\rightarrow B\gamma
\gamma)/M_{A}\approx1.1\times10^{-6}$.}
\label{Fig4}%
\end{figure}

Although the single production of $S_A$ violates the charge $\chi$, 
it is instructive to determine the evolution of $\Lambda$ as a function of $M_A$ 
in this simple scenario (a more realistic model will be discussed below), this is shown in Fig.~\ref{Fig4}.  
At this stage, the three-body scenario does not really help to explain the
largeness of the $\gamma\gamma$ production rate required to match the observed
anomaly, and actually fares worse than the two body scenario because of the
extra phase-space suppression, and of the higher dimensionality of the operators.
For the fermionic and vector
cases, the higher dimension of the operators forces $\Lambda$ to be lower.

To improve the situation, and provide one realistic setting in which the three
body scenario would drive the diphoton anomaly, let us reconsider the decay
and production.

\subsubsection*{Nearly resonant decay}

As a first step, in view of the proximity of $\Lambda$ and $M_{A}-M_{B}$, it
is reasonable to expect that the underlying dynamics could be felt. Consider
for instance the exchange of an off-shell scalar $X$, see Fig.~\ref{Fig0}. Its
impact is to replace two powers of the scale $\Lambda$ by%
\begin{equation}
\frac{1}{\Lambda^{2}}\rightarrow\frac{-1}{m_{\gamma\gamma}^{2}-M_{X}%
^{2}+iM_{X}\Gamma_{X}}\;,\label{BW}%
\end{equation}
in the effective interactions of Eq.~(\ref{All2G}). The scale factor remaining
after this substitution then concerns only the $X\gamma\gamma$ and $XAB$
couplings (the former has mass dimension $-1$, while the latter has mass
dimension $1$, $0$, and $-1$ when $A,B$ are scalars, fermions, or vectors,
respectively). Clearly, when $M_{X}$ is only barely larger than $M_{A}-M_{B}$,
the slightly off-shell $X$ propator strengtens the peak of the differential
rate at high $m_{\gamma\gamma}^{2}$. Further, simply for dimensional reasons,
the constraints on the scale $\Lambda$ tuning the $X\rightarrow\gamma\gamma$
vertex are then much weaker, and tend towards those typical of the two-photon
resonance scenario (see \textit{e.g.} Ref.~\cite{Franceschini:2015kwy}). More
generally, such a conclusion can be reached whenever a form-factor
$F(m_{\gamma\gamma}^{2})$, whose typical expansion would be $F(m_{\gamma
\gamma}^{2})=1+\alpha m_{\gamma\gamma}^{2}/M_{A}^{2}+\mathcal{O}%
(m_{\gamma\gamma}^{4}/M_{A}^{4})$, needs to be inserted to account for the
short-distance dynamics. With $\alpha>0$, such a form-factor further
strengtens the peak of the differential rate at high $m_{\gamma\gamma}^{2}$.

\subsubsection*{Production of the parent resonance}

Given our hypothesis of a new conserved quantum number for the $A$ and $B$
particles, it would be more adequate to consider the $gg\rightarrow A+B$
process. There is indeed no obstruction to replace the photon by the gluon
field strength in the effective operators. In that case, $A\rightarrow
B\gamma\gamma$ would be accompanied by $A\rightarrow Bgg$. The dijet invariant
mass would follow the same distribution as the two-photon one, and would thus
peak again around $750~$GeV. Note, however, that the decay rate $A\rightarrow
Bgg$ and the production mechanism $gg\rightarrow A+B$ are far less easy to
relate than $gg\rightarrow A$ and $A\rightarrow gg$ in the two-body
hypothesis. The typical scale of the $gg\rightarrow A+B$ process is higher
than $M_{A}$ since the gluons must carry at least an energy equal to
$M_{A}+M_{B}$ in their center of mass. Furthermore, since the scale $\Lambda$
cannot be much higher than $M_{A}$, the structure of the effective $ggAB$
vertex is likely to become very relevant for production. It could even begin
to resonate if we imagine that these vertices arise from tree-level or loop
processes, in which case a cascade decay mechanism would need to be
considered. This means that in principle, the $gg\rightarrow A+B$ production
could be quite strong even with a moderate dijet production $A\rightarrow
Bgg$, in agreement with the current absence of a signal in the latter.

To make this statement more precise we should give up our model-independent
formalism. So, for illustration, let us consider a scenario in which both the
effective $A\rightarrow Bgg$ and $A\rightarrow B\gamma\gamma$ decays are
induced by the exchange of an off-shell scalar $X$, with $A$ and $B$ either
scalars, fermions, or vectors. Then, the production of the $A$ particle can
proceed via an on-shell $X$, so that in the narrow width approximation,%
\begin{equation}
\sigma(pp\overset{}{\rightarrow}\gamma\gamma BB)=\frac{C_{gg}(M_{X})}{M_{X}%
s}\Gamma\left(  X\rightarrow gg\right)  \frac{1}{\Gamma_{X}}\Gamma\left(
X\rightarrow AB\right)  \frac{1}{\Gamma_{A}}\Gamma\left(  A\rightarrow
B\gamma\gamma\right)  \;,
\end{equation}
where $C_{gg}(M_{X})$ is the $gg$ partonic integral for a resonance of mass $M_X$ at the $\sqrt{s}=13$ TeV LHC. We evaluated $C_{gg}(M_{X})$ at a scale $\mu_F=M_X$ using the NNPDF 3.0 parton distribution functions~\cite{Ball:2014uwa},
\begin{equation}%
\begin{tabular}
[c]{l|ccccc}%
$M_{X}$\;[TeV] & $0.9$ & $1.0$ & $1.5$ & $2.0$ & $3.0$\\\hline
$C_{gg}(M_{X})$ & $\;863\;$ & $\;\;491\;\;$ & $\;\;47\;\;$ &
$\;\;\;7\;\;\;$ & $\;\;\;\;0.3\;\;\;\;$%
\end{tabular}
\end{equation}
Assuming that BR$\left(  X\rightarrow AB\right)  \approx1$ and setting
$\sigma(pp\overset{}{\rightarrow}\gamma\gamma BB)\approx6$%
~fb~\cite{Ellis:2015oso}, this simplifies to%
\begin{equation}
\frac{\Gamma\left(  X\rightarrow gg\right)  }{M_{X}}\text{BR}\left(A\rightarrow
B\gamma\gamma\right)  =\frac{s\sigma(pp\overset{}{\rightarrow}\gamma\gamma
BB)}{C_{gg}(M_{X})}\approx\frac{2.5\times10^{-3}}{C_{gg}(M_{X})}\;.
\end{equation}
Thus, even though $C_{gg}(M_{X})$ quickly decreases with increasing $M_{X}$,
this does not imply that $\Gamma\left(  A\rightarrow B\gamma\gamma\right)  $
should also increase. If $A\rightarrow B\gamma\gamma$ and $A\rightarrow Bgg$
are the only two available decay modes, then all is needed is a non-suppressed 
BR$\left( A\rightarrow B\gamma\gamma\right)$. Small BR$\left(  A\rightarrow Bgg\right)$
is also preferable in order to suppress the potential di-jet signature.
Taking for definiteness BR$\left( A\rightarrow B\gamma\gamma\right)=1/2$, 
the effective scale is only constrained by the initial $X$ production:
\begin{equation}
\frac{5\times 10^{-3}}{C_{gg}(M_{X})}\approx\frac{\Gamma\left(  X\rightarrow
gg\right)  }{M_{X}}=8\pi\alpha_{3}^{2}\frac{M_{X}^{2}}{\Lambda_{X}^{2}}\;,
\end{equation}
where in the last equality we assumed an effective coupling $(g_{3}%
^{2}/\Lambda_{X})XG_{\mu\nu}G^{\mu\nu}$. Such small rates actually push the
scale to very high values,
\begin{equation}%
\begin{tabular}
[c]{l|ccccc}%
$M_{X}$\;[TeV] & $0.9$ & $1.0$ & $1.5$ & $2.0$ & $3.0$\\\hline
\;$\Lambda_{X}$\;[TeV] & $\;\;184\;\;$ & $\;\;154\;\;$ & $\;\;72\;\;$ &
$\;\;38\;\;$ & $\;\;12\;\;$%
\end{tabular}
\end{equation}
Note that this does not imply too long lifetimes for the $A$ particle, since
the scale $\Lambda_X$ only tunes the $X\gamma\gamma$ and $Xgg$ couplings (in
other words, remember that in Eq.~(\ref{scalar}), four powers of $\Lambda$
should be replaced by $M_{X}^{4}$, see Eq.~(\ref{BW})).

Finally, it is instructive to compare this interpretation with the two-body scenario. If $X$ is directly responsible for the diphoton anomaly, we can write
\begin{equation}
\sigma(pp\overset{}{\rightarrow}\gamma\gamma,jj)=\frac{C_{gg}(M_{X})}%
{s}\frac{\Gamma\left(  X\rightarrow gg\right)  \Gamma\left(  X\rightarrow
\gamma\gamma,gg\right)  }{M_{X}\Gamma_{X}}\;.
\end{equation}
Then, assuming $\Gamma\left(  X\rightarrow gg\right)  \ll\Gamma_{X}\approx\Gamma\left(
X\rightarrow\gamma\gamma\right)$, the scale derived from $X\rightarrow gg$ are consistent with those quoted above. However, reproducing the diphoton production rate forces $\Gamma\left(  X\rightarrow\gamma\gamma\right)$ to be very large.  Assuming an interaction of the form $(e^{2}/\Lambda_{\gamma})XF_{\mu\nu}F^{\mu\nu}$, its effective scale has to be dangerously lower than $\Lambda_{X}$ (see Fig.~1 in Ref.~\cite{Franceschini:2015kwy}). In the three-body scenario on the
contrary, both $\Gamma\left(  X\rightarrow gg\right)  \approx\Gamma\left(
X\rightarrow\gamma\gamma\right)  \ll\Gamma_{X}$ can be tiny because the
large diphoton rate is ensured thanks to the large BR$\left(  X\rightarrow
AB\right)  $ and BR$\left(  A\rightarrow B\gamma\gamma\right)  $. This is
certainly consistent with the dimensions of the interactions: $Xgg$ and
$X\gamma\gamma$ are necessarily suppressed by some high scale $\Lambda_X$, but
the $ABX$ vertex could even be renormalizable when $A$, $B$ are scalars or
fermions. Thus, in the present scenario, the diphoton signal overwhelmingly
arises from the three-body decay of the $A$ particle, while the $gg\rightarrow
X\rightarrow\gamma\gamma,gg$ remains tiny.

\section{Concluding remarks}

Theoretically, interpreting the anomaly observed in the two-photon invariant
mass spectrum as arising from a three-body process $A\rightarrow B\gamma
\gamma$ has two main advantages. There is no need to account for a large width
for the parent particle, and it is quite natural for the involved new
particles to share some conserved quantum numbers. This should be welcome for
many models where such charges are introduced, \textit{e.g.}, to ensure the
stability of a light DM candidate or a suppression of FCNC (as R-parity in
supersymmetry). On the other hand, the main disadvantage of this scenario is
the higher dimensionality of the effective interactions. If taken seriously,
the diphoton anomaly is surprisingly large, and is already non-trivial to
reproduce in the two-body decay scenario. With three bodies, the situation
seems to worsen with the effective interaction scale ending up even lower.
While this is generically true, the three-body nature of the process opens new
alternative production mechanisms and there are ways to circumvent this
problem. We have provided one such example, in which the $AB\gamma\gamma$ and
$ABgg$ interactions arise from the exchange of a scalar resonance $X$. 
In that case, the scale of the New Physics inducing both the
$X\gamma\gamma$ and $Xgg$ vertices could still be far above the TeV. This is
actually an improvement over the pure 750~GeV diphoton resonance scenario, in
which the scale of at least one of these vertices has to be close to the TeV,
see \textit{e.g.} Ref.~\cite{Franceschini:2015kwy}.

Experimentally, discriminating this scenario from a pure two-photon resonance
would of course be achieved with a better resolution of $d\Gamma
(pp\rightarrow\gamma\gamma)/dm_{\gamma\gamma}^{2}$, especially below the
750~GeV peak. Even if modulated by some effective form
factor, the shape of the differential rate should significantly depart from
the simple Breit-Wigner expected for a diphoton resonance. On the other hand,
we find that the shape of the differential rate does not constrain the spin of
the $A,B$ particles, with scalars, fermions, or vectors producing essentially
the same signature.

To further confirm the three-body nature of the process, the presence of some
missing energy could be a tell-tale sign since the daughter particle $B$ in
$A\rightarrow B\gamma\gamma$ is never strictly at rest (the differential rate
vanishes at the end-points). For the same reason,
the two photons are never flying precisely back-to-back in the $A$ rest frame,
so their angular distribution could provide complementary information. Note
in addition that missing energy may arise if $B$ particle also accompany the
production of the $A$ particle, like in a $gg\rightarrow B+[A\rightarrow
B\gamma\gamma]$ chain. 

On the other hand, $Z\gamma$ and $ZZ$ signals, which should be seen at some
point, would not help pinpoint the nature of the process since they should
occur in the same ratios respective to $\gamma\gamma$ as in the two-body decay
scenario. Similarly, some peak in the dijet spectrum would seem likely but
would not be characteristic. Note however that this depends on the true
production mechanism for the parent particle, which could proceed instead
through some cascades from other yet unknown particles. Finally, lepton pairs
or quark pairs would be generically suppressed in this scenario, and no signal
should be seen there.

\section*{Acknowledgments}

We are grateful to Sabine Kraml for numerous discussions. This work has been
supported in part by the ``Investissements d'avenir, Labex ENIGMASS''.

\end{document}